\documentclass[aps,prl,reprint,groupedaddress,nofootinbib]{revtex4-1}
\usepackage{graphicx, color, soul}
\usepackage{siunitx}
\usepackage{amsmath}
\usepackage{braket}
\usepackage{color,soul,url}
\usepackage{float}

\begin{document}
\title{Point-Defect-Localized Bound States in the Continuum in Photonic Crystals and Structured Fibers}
\author{Sachin Vaidya,$^{1,*}$ Wladimir A. Benalcazar,$^1$ Alexander Cerjan,$^{1, 2, 3}$ and Mikael C. Rechtsman$^1$}
\affiliation{
 $^1$Department of Physics, The Pennsylvania State University, University Park, Pennsylvania 16802, USA\\
 $^2$Sandia National Laboratories, Albuquerque, New Mexico 87123,  USA\\
 $^3$Center for Integrated Nanotechnologies, Sandia National Laboratories, Albuquerque 87123, New Mexico, USA
}

\def\thefootnote{*}\footnotetext{sxv221@psu.edu}

\date{\today}

\begin{abstract}
We show that point defects in two-dimensional photonic crystals can support bound states in the continuum (BICs). The mechanism of confinement is a symmetry mismatch between the defect mode and the Bloch modes of the photonic crystal. These BICs occur in the absence of bandgaps and therefore provide an alternative mechanism to confine light. Furthermore, we show that such BICs can propagate in a fiber geometry and exhibit arbitrarily small group velocity which could serve as a platform for enhancing non-linear effects and light-matter interactions in structured fibers.
\end{abstract}
\maketitle

Over the last three decades, photonic crystals (PhCs) have been shown to exhibit exceptional confinement and transport properties that exploit the existence of a photonic bandgap, a band of frequencies where no electromagnetic waves may propagate \cite{photoniccrystalsbook, photoniccrystalsbook2, PhC1, PhC2}. Photonic bandgaps can inhibit spontaneous emission of embedded quantum emitters \cite{PhC_SE_1, PhC_SE_2, PhC_SE_3, PhC_SE_4}, facilitate slow-light through band-edge operation \cite{SlowLightReview} or host localized defect modes that can serve as high-Q resonators or waveguides. Confined defect modes form the basis of many devices such as PhC fibers \cite{PhCFiber_Knight, PhCFiber_Russell}, spectral filters, and lasers \cite{Laser, Laser2} and to achieve near-perfect confinement, defect modes are constructed to lie within photonic bandgaps so as to spectrally isolate them from the extended states of the PhC. However, this necessitates the use of materials with a sufficiently high refractive index to open complete gaps. An alternative mechanism for confinement could circumvent the need for bandgaps, enabling the use of many low-refractive index materials such as glasses and polymers as well as increasing design flexibility for the realization of PhC-based devices.

One possible way to achieve this is by using bound states in the continuum (BICs). BICs are eigenmodes of a system that, despite being degenerate with a continuum of extended states, stay confined -- this confinement may result from a variety of mechanisms \cite{BICsreview}. For example, modes of a PhC slab that lie above the light line of vacuum and therefore could radiate, can remain perfectly bound to the slab \cite{BICs1, BICs2, BICs3, BICs4, BICs5, BICs6}. Previous designs with BICs have mostly shown confinement of a mode in one dimension lower than that of the environment. Recently, corner-localized BICs were predicted and observed in two-dimensional chiral-symmetric systems with higher-order topology \cite{HOTIBICs1, HOTIBICs2}. However, chiral (sub-lattice) symmetry is, in general, strongly broken in all-dielectric PhCs. Indeed, confinement in the continuum has to this point not yet been achieved in point defects embedded inside multi-dimensional PhCs.

In this work, we predict the existence of BICs that are exponentially confined to point defects in a two-dimensional PhC environment. The defect cavity and bulk PhC are designed such that radiation leakage is prohibited due to a symmetry mismatch between the defect mode and the ambient continuum states. The BICs proposed here are protected by the simultaneous presence of time-reversal symmetry (TRS) and the point group of the lattice and as such are robust as long as these symmetries are maintained. As an application for these BICs, we also show how they can circumvent bandgap requirements and be used as propagating fiber modes with arbitrarily small group velocity in a low-contrast slow-light PhC fiber.

\begin{figure}
\centering
\includegraphics[scale = 0.215]{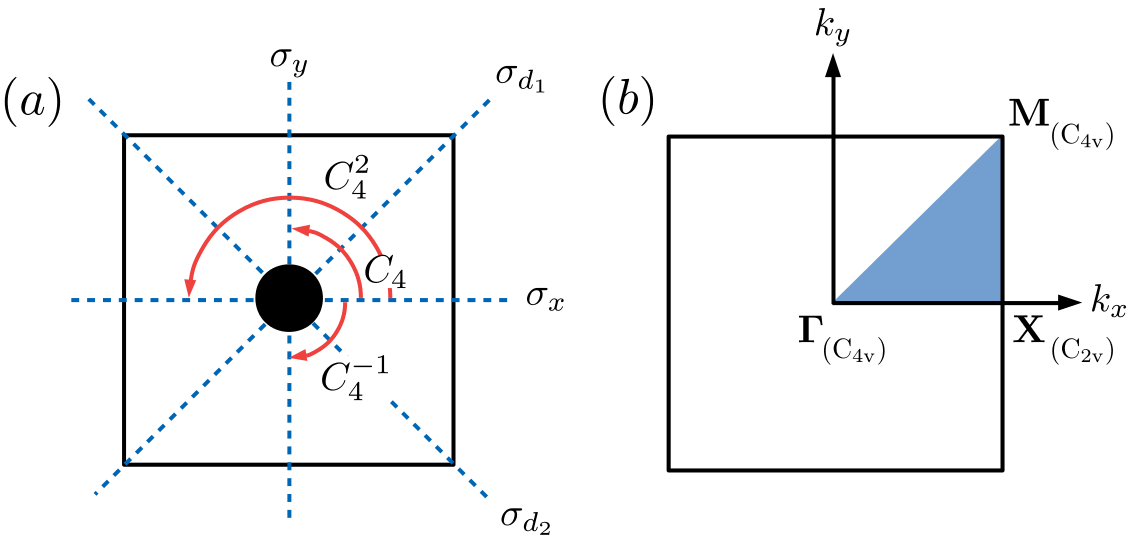}
\caption{(a) The unit cell of a two-dimensional PhC consisting of circular discs. The symmetry operators of the $C_{4v}$ point group are labelled. (b) The Brillouin zone of the PhC showing its HSPs and the little groups under which the HSPs are invariant. The solid color consists of all momenta that lie within the irreducible Brillouin zone.}
\label{fig:figure1}
\end{figure}

We draw a distinction between our BICs and the previously reported defect modes degenerate with Dirac points in 2D PhCs \cite{dirac1, dirac2, dirac3, diracexp1, diracexp2, diracexp3}. In the latter case, the confinement of light to a defect site is due to a vanishing density of states at the Dirac point, which is where that confined mode's frequency lies. Characteristically, such defect modes exhibit weak confinement due to the algebraic mode profile away from the defect site. In contrast, the defect modes presented here are bona fide symmetry protected BICs that are exponentially localized to the defect site.

 Consider a two-dimensional PhC consisting of a square lattice of discs with dielectric constant $\varepsilon$ and radius $r$ embedded in vacuum. This PhC, as shown in Fig. 1 (a) is invariant under 90$^{\circ}$ rotations ($C_4$, $C_4^2$, $C_4^{-1}$), and reflections along the $x$, $y$ axes and two diagonals ($\sigma_x$, $\sigma_y$, $\sigma_{d_1}$, $\sigma_{d_2}$). These symmetry operations constitute the $C_{4v}$ point group. The irreducible Brillouin zone of this lattice contains three inequivalent high symmetry points (HSPs), namely, $\mathbf{\Gamma} = (0,0)$, $\mathrm{\mathbf{X}} = (\pi/a, 0)$ and $\mathrm{\mathbf{M}} = (\pi/a,\pi/a)$, as shown in Fig. 1 (b). The HSPs $\mathbf{\Gamma}$ and $\mathrm{\mathbf{M}}$ are invariant under the full $C_{4v}$ group, while $\mathrm{\mathbf{X}}$ is invariant only under the little group, $C_{2v}$. Eigenmodes of the PhC at a HSP transform according to the irreducible symmetry representations (irrep) of the group under which the HSP is invariant. The $\mathrm{\mathbf{X}}$ point has four possible one-dimensional irreps $(a_1,a_2,b_1,b_2)$ with character table as shown in Table I. Similarly, the $\mathbf{\Gamma}$ and $\mathrm{\mathbf{M}}$ points have four one-dimensional irreps $(A_1,A_2,B_1,B_2)$ and one two-dimensional irrep $(E)$ with character table as shown in Table II \cite{photoniccrystalsbook2}. The eigenmodes of a $C_{4v}$ symmetric PhC that transform according to the two-dimensional irrep $(E)$ of the $C_{4v}$ point group, commonly manifest as quadratic two-fold degeneracies at $\mathbf{\Gamma}$ and $\mathrm{\mathbf{M}}$ in the presence of TRS. When $C_{4v}$ is broken, this degeneracy splits into two Dirac points as long as inversion and TRS are retained. However, breaking TRS can lift the degeneracy completely \cite{EffectivetheoryofQD}.

\begin{figure*}
\centering
\includegraphics[scale = 0.32]{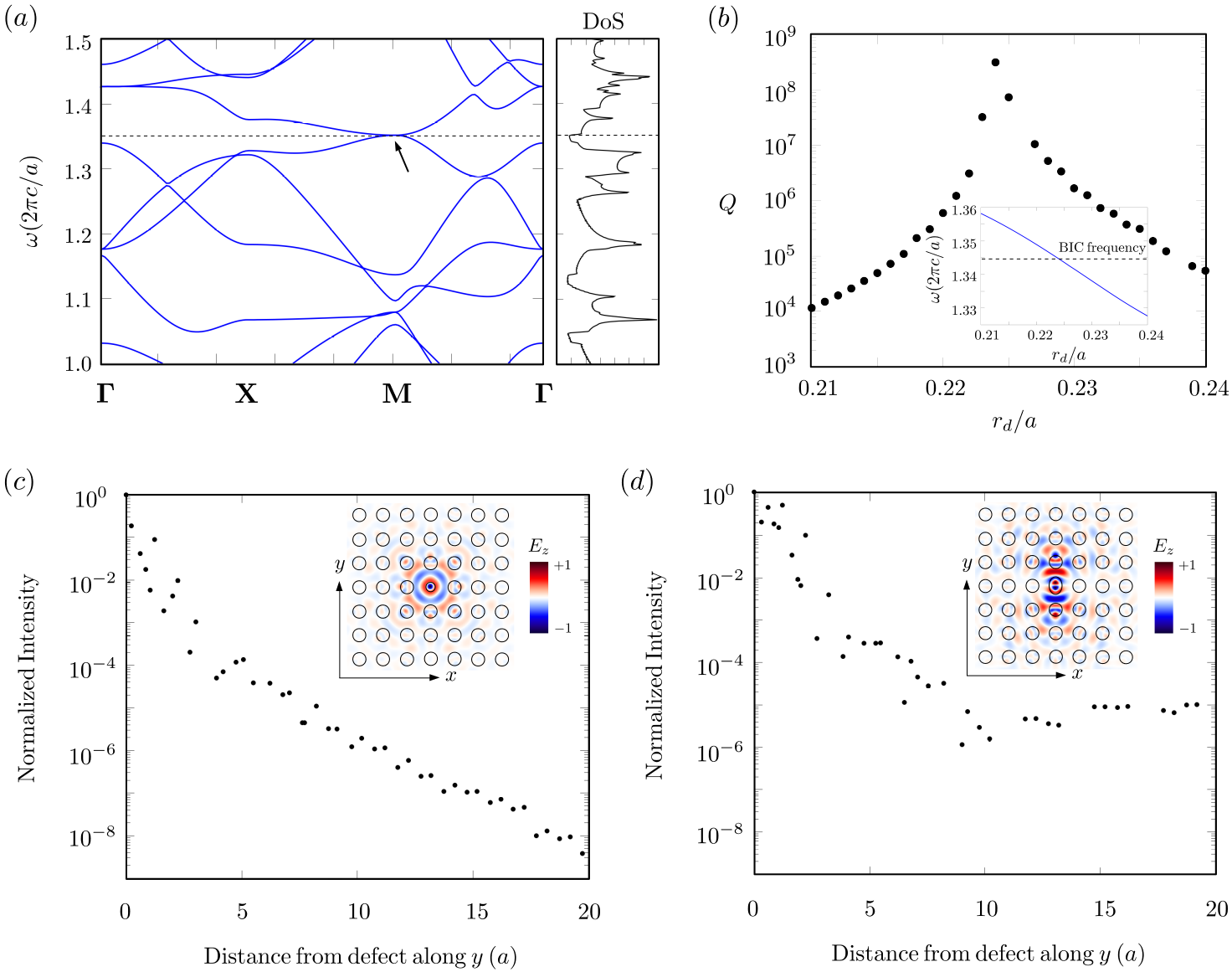}
\caption{(a) The TM bands and photonic DoS of a square lattice of dielectric discs of $\varepsilon = 4$ and $r/a = 0.275$ calculated using MPB \cite{MPB}. The spectrally-isolated two-fold degeneracy is marked with an arrow. (b) Quality factor (Q) of the defect mode as a function of defect radius $(r_d)$. The sharp divergence in Q indicates the existence of a BIC at $r_d/a = 0.224$. The inset shows the dependence of the defect mode frequency on $r_d$. (c) The $\mathbf{E}$-field intensity envelope of the BIC showing exponential localization as a function of distance (along the $y$-axis) from the defect site. The inset shows the $z$-component of the $\mathbf{E}$-field of the BIC, extracted from FDTD simulations.  (d) The $\mathbf{E}$-field intensity envelope of the resonance when the symmetry of the supercell is reduced from $C_{4v}$ to $C_{2v}$. The inset shows the $z$-component of the $\mathbf{E}$-field of the resonance, extracted from FDTD simulations.} 

\label{fig:figure2}
\end{figure*} 
 
\begin{table}[h]
\centering
\setlength{\tabcolsep}{8pt}
\begin{tabular}{ c|c c c c} 
 \hline
  $C_{2v}$ & $\mathrm{I}$ & $C_2$ & $\sigma_x$ & $\sigma_y$ \\ 
 \hline
 $a_1$ & \hskip 6.75 pt 1 & \hskip 6.75 pt 1 & \hskip 6.75 pt 1 & \hskip 6.75 pt 1\\
 $a_2$ & \hskip 6.75 pt 1 & \hskip 6.75 pt 1 & $-1$ & $-1$\\
 $b_1$ & \hskip 6.75 pt 1 & $-1$ & \hskip 6.75 pt 1 & $-1$\\
 $b_2$ & \hskip 6.75 pt 1 & $-1$ & $-1$ & \hskip 6.75 pt 1\\
 \hline
\end{tabular}
	\caption{Character table for the $C_{2v}$ point group.}
	\label{tab:C2v}
\vskip 12 pt
\begin{tabular}{ c|c c c c c } 
 \hline
  $C_{4v}$ & $\mathrm{I}$ & $2C_4$ & $C_2$ & $2\sigma_{x,y}$ & $2\sigma_{d_1,d_2}$ \\ 
 \hline
 $A_1$ & \hskip 6.75 pt 1 & \hskip 6.75 pt 1 & \hskip 6.75 pt 1 & \hskip 6.75 pt 1 & \hskip 6.75 pt 1\\
 $A_2$ & \hskip 6.75 pt 1 & \hskip 6.75 pt 1 & \hskip 6.75 pt 1 & $-1$ & $-1$\\
 $B_1$ & \hskip 6.75 pt 1 & $-1$ & \hskip 6.75 pt 1 & \hskip 6.75 pt 1 & $-1$\\
 $B_2$ & \hskip 6.75 pt 1 & $-1$ & \hskip 6.75 pt 1 & $-1$ & \hskip 6.75 pt 1\\
 $E$ & \hskip 6.75 pt 2 & \hskip 6.75 pt 0 & $-2$ & \hskip 6.75 pt 0 & \hskip 6.75 pt 0\\
 \hline
\end{tabular}
	\caption{Character table for the $C_{4v}$ point group.}
	\label{tab:C4v}
\end{table}

We now describe the general mechanism for creating defect-localized BICs. By changing the geometric parameters of the lattice, the band dispersion of a $C_{4v}$ and TRS-symmetric PhC can be designed such that the two-fold degeneracy at either $\mathbf{\Gamma}$ or $\mathrm{\mathbf{M}}$ is spectrally isolated from other bands. In a large system consisting of many unit cells of such a PhC (a supercell), a single defect site with radius $r_d \ne r$ is introduced at the center. This creates modes with a significant support on the defect site that generally radiate by hybridizing with the bulk states of the PhC, forming leaky resonances that are characterized by a complex frequency with a negative imaginary part. The frequency of such modes can be tuned by changing the parameters of the defect site such as size or dielectric constant. When the real part of the frequency of the defect mode exactly matches that of the spectrally-isolated two-fold degeneracy of the bulk, it becomes a perfectly confined BIC provided that the defect mode transforms according to a one-dimensional irrep that is orthogonal to the two-dimensional irrep of the bulk. The presence of this BIC can be inferred from the vanishing of the imaginary part of the frequency and hence a diverging quality factor, $Q = -\mathrm{Re}(\omega)/2\mathrm{Im}(\omega)$, of the defect mode.

To demonstrate this, we simulate this system using finite-difference time domain method (FDTD) as implemented in MEEP \cite{MEEP}. The bulk band requirements are easily met in a simple square lattice of discs with dielectric constant $\varepsilon = 4$ and radius $r/a = 0.275$, where $a$ is the lattice constant in both $x$ and $y$ directions. The chosen values of $\varepsilon$ and $r/a$ allow the spectrally-isolated two-fold degeneracy to occur between TM bands 10 and 11 at the $\mathbf{M}$ point as shown in Fig. 2 (a). The photonic density of states (DoS), also shown in the same figure, is given by $\mathrm{DoS}(\omega) = \sum_n \int_{\mathbf{k}\in \mathrm{BZ}} \delta[\omega-\omega_{n}(\mathbf{k})] \mathrm{d}\mathbf{k}$, where $\omega_n(\mathbf{k})$ is the frequency eigenvalue at the momentum $\mathbf{k}$ and band index $n$. Since each band undergoes an extremum at the degeneracy, the DoS exhibits a jump-discontinuity-type Van Hove singularity between two finite and non-zero values. The non-vanishing set of states at the degeneracy forms the continuum within which a BIC can be created.

In a large supercell, we now introduce a defect by changing the radius ($r_d \ne r$) of a single disc in the center of the supercell. As we scan the values of $r_d$, a BIC emerges for the specific value of the defect radius that corresponding to a mode with the exact frequency of the bulk degeneracy. This is seen from the sharp divergence of the Q-factor of the defect mode as shown in Fig. 2 (b). Examining the mode profile shown in the inset of Fig. 2 (c) reveals that the defect mode transforms according to the irrep $A_1$ which is prevented from mixing with the basis modes of the orthogonal two-dimensional irrep, $E$, of the bulk. Moreover, the mode shows very strong exponential localization to the defect site which can be seen by plotting the intensity envelope as shown in Fig. 2 (c). Another important feature of this BIC is its occurrence above $\omega a/2\pi c = 1$. This implies that the lattice constant of the bulk PhC is larger than the wavelength of the BIC mode, a property which could prove useful for fabrication, because features sizes would need not be subwavelength.

\begin{figure*}[t]
\centering
\includegraphics[scale = 0.31]{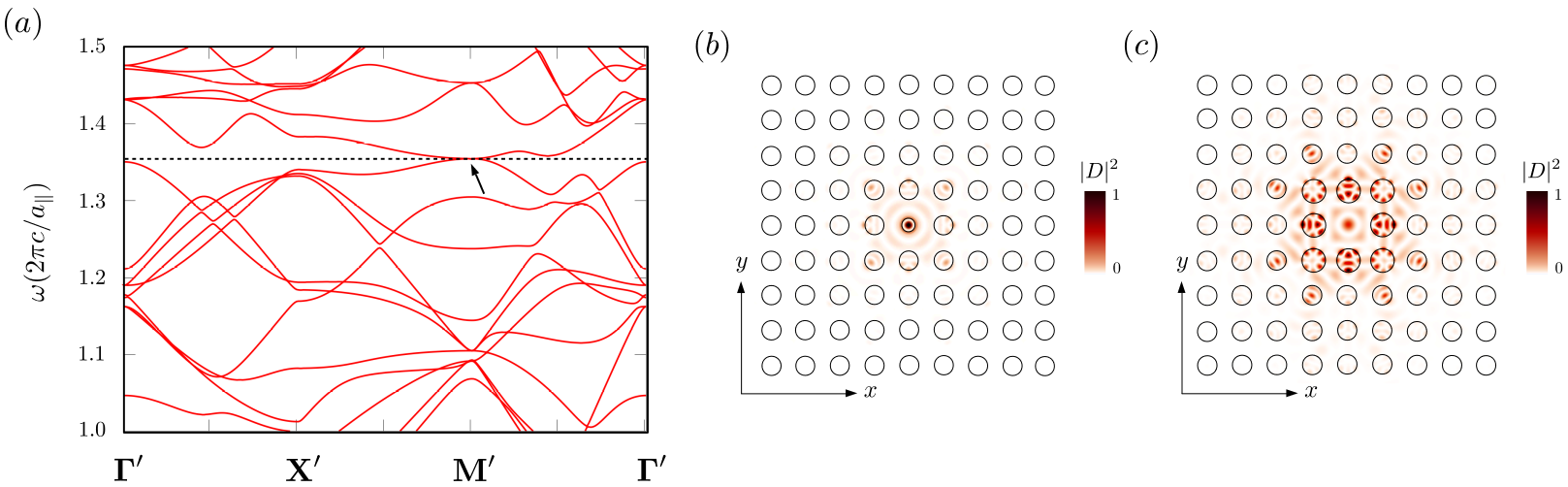}
\caption{(a) The $k_{\parallel}$-band structure of the defect-free PhC fiber at $k_z = 0.18$  $(2 \pi/a_{\parallel})$. The spectrally-isolated two-fold degeneracy is marked with an arrow. (b) $\mathbf{D}$-field intensity profile of a solid-core fiber BIC mode that occurs at $k_z = 0.18$  $(2 \pi/a_{\parallel})$. (c) $\mathbf{D}$-field intensity profile of a hollow-core-like fiber BIC mode. }
\label{fig:figure3}
\end{figure*}

To conclusively show that this BIC is indeed symmetry protected, we change the defect site from a disc to a filled ellipse, which reduces the symmetry of the supercell from $C_{4v}$ to $C_{2v}$. Due to this deformation, the degeneracy between the two modes that formed the two-dimensional irrep, $E$, of $C_{4v}$ is very slightly lifted and the resultant non-degenerate modes have the one-dimensional irreps $b_1$ and $b_2$ of $C_{2v}$. As before, we vary the defect size to tune the frequency of the defect mode and find a maximum $Q \sim 10^4$ indicating that the mode is not a BIC but a resonance. Indeed, the field pattern of the defect mode as shown in the inset of Fig 2 (d), transforms according to $b_2$, which coincides with one of the irreps of the bulk enabling the defect and bulk modes to couple and create a leaky resonance with a finite Q (see Supplementary Material). This is also evident from the intensity envelope of the resonance as shown in Fig 2 (d) that markedly demonstrates the lack of exponential confinement to the defect site. Displacing the defect site away from the center also breaks the $C_{4v}$ symmetry of the supercell and has a similar effect of degrading the Q-factor of the mode (see Supplementary Material).

The symmetry mismatch between the defect mode and bulk bands requires the existence of a spectrally-isolated two-fold degeneracy in the bulk PhC so the question naturally arises: how easy is it to design this bulk band requirement? It is clear from our findings that even simple PhC designs are able to satisfy the requirements for reasonably low dielectric contrast and in fact, the feature in the TM bands of the PhC discussed in Fig. 2 (a) persists down to $\varepsilon = 3$ for a slightly smaller value of $r/a$. Furthermore, such quadratic degeneracies can also occur at the $\mathbf{\Gamma}$ point in $C_{3v}$ and $C_{6v}$ symmetric lattices, forming two-dimensional irreps of the respective point groups. In the Supplementary Material, we outline a method for finding optimized structures with tunable parameters that exhibit such degeneracies.

For traditional defect modes in 2D PhCs, it suffices to have a bandgap for one polarization, either TE or TM, since they constitute orthogonal subspaces that do not mix. However, for applications such as PhC fibers, (i.e., where the 2D pattern described above is extruded in the third direction, $z$, and $k_z \ne 0$ generally), the distinction between TE and TM is lost and one requires an overlapping bandgap for both polarizations to confine defect modes. In particular, slow-light PhC fibers rely on the existence of a complete bandgap at $k_z=0$ which persists for a small range of $k_z$ \cite{Slowlight1, slowlight2, slowlight3}. The arbitrarily small group velocity of the propagating modes in such fibers is achieved by operating near the $k_z=0$ band edge. These slowly-propagating modes can then be used to strongly enhance interactions of light with either the dielectric material itself or an infiltrated material \cite{RbHollowCore, SemiconductorInfil}, depending on whether the fiber hosts a solid or hollow core. Thus, the design of these fibers requires a high dielectric contrast to open a complete bandgap at $k_z=0$. To the best of our knowledge, the smallest contrast for which a complete bandgap exists for 2D PhCs is for $\varepsilon = 4.41$ \cite{supercellbandgaps}. We now extend the idea of point-defect-localized BICs to propagating slow-light fiber modes, circumventing the requirement for a complete bandgap. 

The fiber design that we propose is identical to an extruded version of the 2D PhC discussed before, now consisting of cylinders extended along the direction of propagation in the fiber. However, since the distinction between TE and TM polarizations is lost, the spectrally-isolated two-fold degeneracy of the bulk must occur in the full band structure in order to create a BIC. This is easily achieved in our structure for a range of $k_z$ values around $0$. For instance, Fig. 3 (a) shows the band structure of the fiber with $\varepsilon = 4$, $r/a = 0.2755$ at $k_z = 0.18$ $(2\pi/a_{\parallel})$, where $a_{\parallel}$ is the lattice constant in the $x,y$ plane. As before, we introduce a defect site and tune the radius $r_d$ and find a BIC at $r_d/a = 0.230$ for this particular value of $k_z$. The field profile of the BIC is plotted in Fig. 3 (b), forming a solid-core mode and displaying strong confinement to the defect site. Since the spectrally-isolated two-fold degeneracy persists down to $k_z = 0$, the group velocity of this BIC along the length of the fiber $(v_{g_z} = \mathrm{d}\omega / \mathrm{d} k_z)$ can be made arbitrarily small with an appropriate choice of $r_d$. It is also possible to create a hollow-core-like fiber mode where the BIC has reasonable support in the air region. To achieve this, we omit the central defect site and instead tune the radius of the nearest eight sites uniformly so as to maintain $C_{4v}$ and find a BIC as shown in Fig. 3 (c).

The BICs presented here could be experimentally realized in a variety of systems. For example, these principles could be applied to create high-Q nanocavities in gapless PhC slabs where some vertical leakage is unavoidable but in-plane leakage could be suppressed through the symmetry mismatch mechanism. Functionally, these modes would behave similarly to run-of-the-mill PhC slab-based cavities that rely on a bandgap but could be realizable in alternative structures with potentially lower dielectric contrast. Similarly, the PhC fiber design discussed here could be implemented straightforwardly by complex fiber drawing techniques \cite{PhCFiber_Russell}. Furthermore, such isolated degeneracies are also known to occur in 3D PhCs which could lead to true gapless confinement of light in all directions such as in structures that are precursors to ones with Weyl points \cite{LingLuDG, 3DPhC1, 3DPhC2, Vaidya_Weyl}. Evidently, these BICs rely solely on symmetry considerations and can also be readily realized using other periodic systems such as acoustic crystals, waveguides \cite{BICs_waveguides, HOTIBICs2} and coupled resonator arrays.

In conclusion, we have proposed BICs that are exponentially localized to defects beyond bandgaps in both 2D PhCs and structured fibers. The PhC slow-light fiber implementation relaxes the need for bandgaps at $k_z=0$ and thus allows for a wider range of materials to be used for their implementation. The results presented here have consequences for the general design of PhC-based devices since the requirement for finding bandgaps could potentially be replaced with finding isolated degeneracies at HSPs, which occur more commonly, at lower dielectric contrast and at higher frequencies in the band structure. Furthermore, it may be possible to use the BIC mechanism to realize hinge modes in higher order photonic topological insulators \cite{HOTI1, HOTI2, HOTI3, HOTIBICs1, HOTIBICs2} due to their structural similarity with PhC fiber modes.

\begin{acknowledgements}
M. C. R. acknowledges support from the Office of Naval Research (ONR) Multidisciplinary University Research Initiative (MURI) grant N00014-20-1-2325 on Robust Photonic Matertials with High-Order Topological Protection as well as the Packard Foundation under fellowship number 2017-66821. W.A.B. is grateful for the support of the Eberly Postdoctoral Fellowship at the Pennsylvania State University. This work was performed, in part, at the Center for Integrated Nanotechnologies, an Office of Science User Facility operated for the U.S. Department of Energy (DOE) Office of Science. Sandia National Laboratories is a multimission laboratory managed and operated by National Technology \& Engineering Solutions of Sandia, LLC, a wholly owned subsidiary of Honeywell International, Inc., for the U.S. DOE’s National Nuclear Security Administration under contract DE-NA-0003525. The views expressed in the article do not necessarily represent the views of the U.S. DOE or the United States Government. Computations for this research were performed on the  Pennsylvania State University’s Institute for Computational and Data Sciences Roar supercomputer.
\end{acknowledgements}

\bibliography{defectBICs}

\end{document}


\onecolumngrid
\raggedbottom
\begin{center}
    \begin{large}
    \textbf{Supplementary Material for \\ Point-Defect-Localized Bound States in the Continuum in Photonic Crystals and Structured Fibers \vskip 12 pt}
    \end{large}
    Sachin Vaidya$^1$, Wladimir A. Benalcazar$^1$, Alexander Cerjan$^{1, 2, 3}$, Mikael C. Rechtsman$^1$\\
    \emph{$^1$Department of Physics, The Pennsylvania State University, University Park, Pennsylvania 16802, USA\\
 $^2$Sandia National Laboratories, Albuquerque, New Mexico 87123,  USA\\
 $^3$Center for Integrated Nanotechnologies, Sandia National Laboratories, Albuquerque 87123, New Mexico, USA}
\end{center}

\section{Q-factor for $\mathbf{C_{4v}}$ and $\mathbf{C_{2v}}$ supercells}

In the main text, we argue that changing the symmetry of the supercell from $C_{4v}$ to $C_{2v}$ allows the defect mode to form a leaky resonance due to a change in the representation of the bulk degeneracy from $E$ to $b_1 \oplus b_2$. This symmetry breaking is achieved by changing the defect site from a circular disc to an elliptical disc with semi-major and semi-minor axes lengths of $0.53a \times t_d$ and $0.4a \times t_d$ respectively where $a$ is the lattice constant and $t_d$ is a tuning parameter. Fig. S1 shows a comparison of quality factors of the defect mode for the two cases and clearly shows the lack of divergence of Q in the $C_{2v}$ symmetric supercell, indicating that the defect mode is indeed a resonance.

\begin{figure}[H]
\centering
\includegraphics[scale = 0.35]{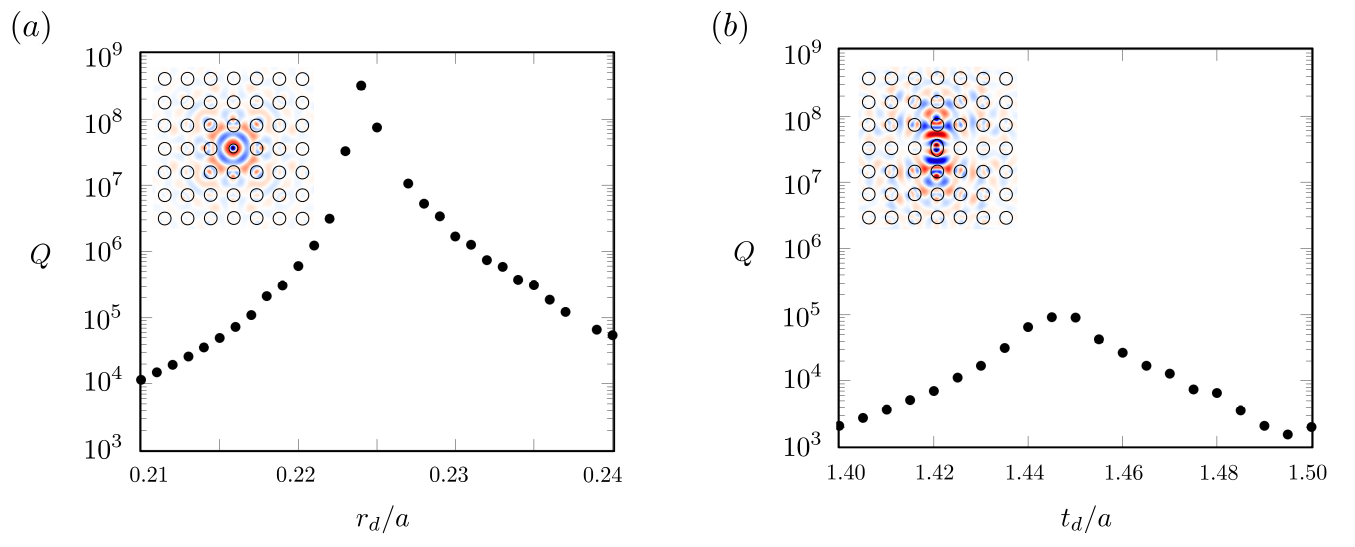}
\caption{(a) Quality factor of the defect mode (irrep $A_1$) in a supercell with $C_{4v}$ symmetry. The divergence in Q shows the appearance of a BIC. (b) Quality factor of a defect mode (irrep $b_2$) in a supercell with only $C_{2v}$ symmetry. The lack of divergence indicates that the defect mode is a resonance. The insets show the defect mode profiles for parameter values corresponding to the maximum Q.}
\label{fig:figure1s}
\end{figure}

\section{Q-factor for shifted defect site}
To assess the impact of symmetry breaking due to fabrication imperfections, we consider the same 2D PhC system as discussed in Fig. 2 of the main text. We then calculate the Q-factor of the defect mode from FDTD simulations where the defect site is displaced along the x-direction as shown in Fig. S4 (a). Fig. S4 (b) shows a plot of the Q-factor as a function of the displacement, $\Delta x$, normalized to the wavelength of the defect mode, $\lambda_d$. As is expected for any symmetry-protected BIC, this perturbation degrades the Q-factor of the BIC. However, we can see that the defect mode still exhibits high Q-factors for typical PhC slab and fiber fabrication errors, which are much less than $\lambda_d/10$. Therefore, any perturbations that are much smaller than the scale of the wavelength will still allow the system to exhibit an ultra-high-Q resonance which includes bends and compressions in the fiber or positional errors in the lattice.
\begin{figure}[H]
\centering
\includegraphics[scale = 0.325]{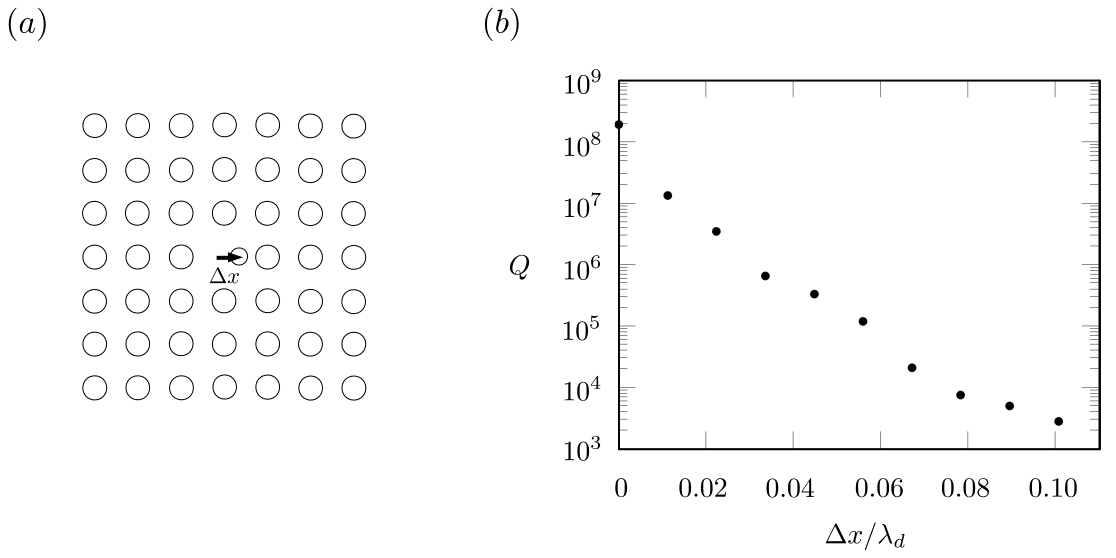}
\caption{(a) Displaced defect site inducing a symmetry breaking in the supercell. (b) Quality factor of the defect mode as a function of the normalized displacement of the defect site.}
\label{fig:figure4s}
\end{figure}

\section{Finding Spectrally Isolated Degeneracies in PhCs using MPB}
The BICs presented in this work can only occur when the bulk band structure exhibits a spectrally-isolated quadratic two-fold degeneracy. Such degeneracies can be found either at $\mathbf{\Gamma}$ or $\mathbf{M}$ in $C_{4v}$ symmetric PhCs or at $\mathbf{\Gamma}$ in $C_{3v}$ and $C_{6v}$ symmetric PhCs. The software package MPB \cite{MPB} outputs the bandgap along a given trajectory in $\mathbf{k}$-space and provides optimization routines to find bandgaps given some free parameters. Here we describe a method to find isolated degeneracies based on the use of this function. The idea is based on the fact that in a PhC where spectrally-isolated degeneracies occur at HSPs, simply detuning away from HSPs by a small amount $\Delta \mathbf{k}$ results in the opening of a small stop band proportional to $\Delta \mathbf{k}^2$. The structural parameters of the PhC can then be optimized to find these stop bands. 

To demonstrate this, we consider the PhC shown in Fig. S2 (a) which consists of three circular discs of radii $r_1$, $r_2$ and $r_3$. The dielectric constant of the high-index material (gray) is $\varepsilon = 2.8$ and that of the low-index material (white) is $\varepsilon = 1$. Using MPB, we run an optimization function on the radii to find the aforementioned stop bands by computing the band structure along the path $(\mathbf{\Gamma}+\Delta \mathbf{k_1})\rightarrow \mathbf{X}  \rightarrow (\mathbf{M}+\Delta \mathbf{k_2}) \rightarrow (\mathbf{\Gamma}+\Delta \mathbf{k_1})$ for some small $\Delta \mathbf{k_1}$, $\Delta \mathbf{k_2}$. A stop band along the detuned path and hence the required degeneracy is found between TM bands 7 and 8 at $\mathbf{\Gamma}$ for $r_1/a = 0.0924$, $r_2/a = 0.4066$ and $r_3/a = 0.4238$, as shown in Fig. S2 (b).

\begin{figure}[H]
\centering
\includegraphics[scale = 0.21]{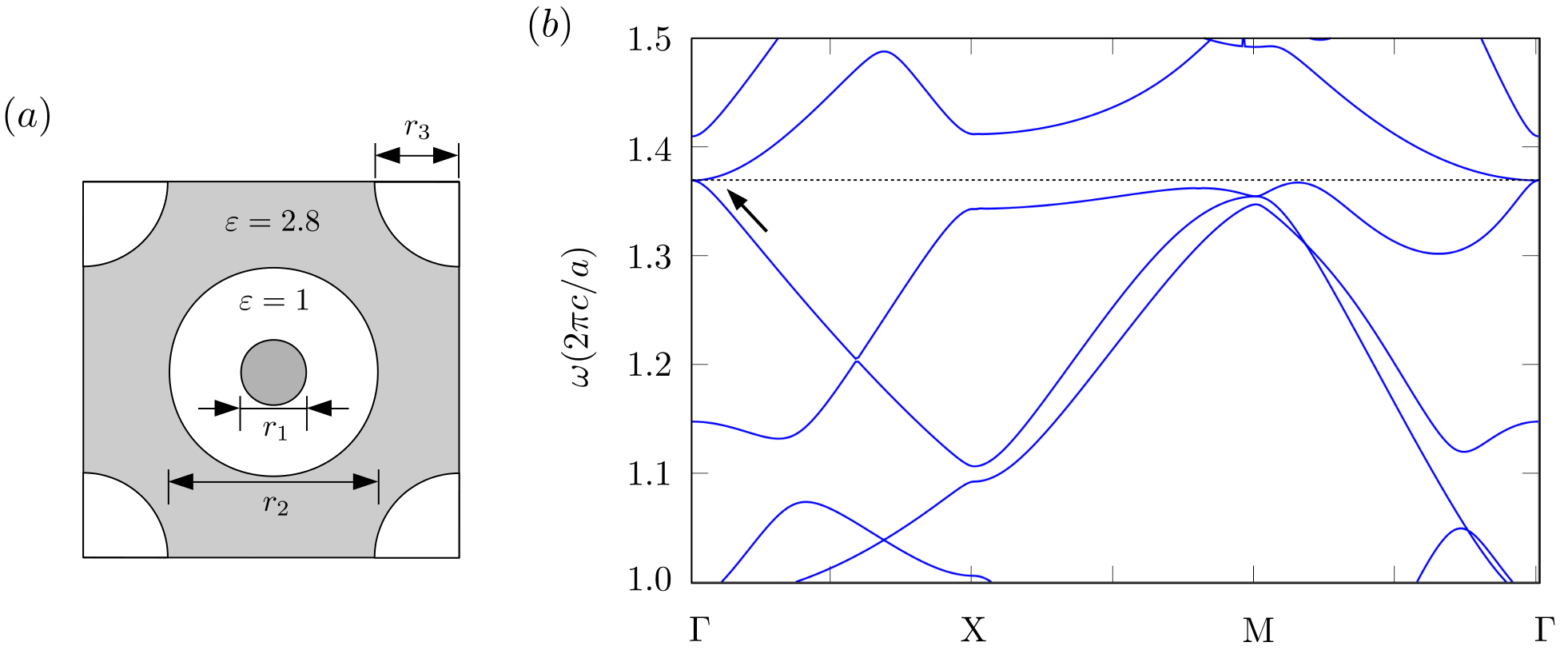}
\caption{(a) The PhC design with three parameters: $r_1$, $r_2$, and $r_3$ made out of a dielectric material with $\varepsilon = 2.8$ (b) TM bands of the PhC shown in (a) for optimized values $r_1/a = 0.0924$, $r_2/a = 0.4066$ and $r_3/a = 0.4238$. The spectrally isolated degeneracy occurs at $\mathbf{\Gamma}$ and is marked with an arrow. }
\label{fig:figure3s}
\end{figure}

\section{Supercell Band Structures for Identifying Defect Modes}
Besides using FDTD, a second way to identify the presence and symmetries of the defect modes is by examining the band structure of a reasonably sized supercell of a PhC with periodic boundaries that contains a defect. To illustrate this, we consider the TM modes of a 2D PhC made of circular discs with $r=0.15$ and $\varepsilon = 6$. This PhC exhibits the spectrally isolated degeneracy between its second and third TM bands. We introduce a defect in this supercell by detuning the radius of the central disc and plot the band structure of this supercell going through the high symmetry points of its small Brillouin zone. Fig. S3 (a) shows that the bulk degeneracy can still be clearly seen at the new $\mathbf{M}$ point in the folded band structure. The defect modes are easily identified by characteristic flat bands in the dispersion of this supercell. Moreover, as the defect radius is varied, the frequencies of the bulk states which have support on all sites of the supercell are barely affected but the frequencies of the defect localized modes are strongly affected. We can then see the effect of tuning the defect size in middle panel of Fig. S3 (a) where the symmetry mismatch between the bulk and defect modes allows for a fine-tuned three-fold degeneracy to occur. The absence of any avoided crossings indicates the formation of a BIC due to a lack of mixing between the bulk and defect modes.

\begin{figure}[H]
\centering
\includegraphics[scale = 0.19]{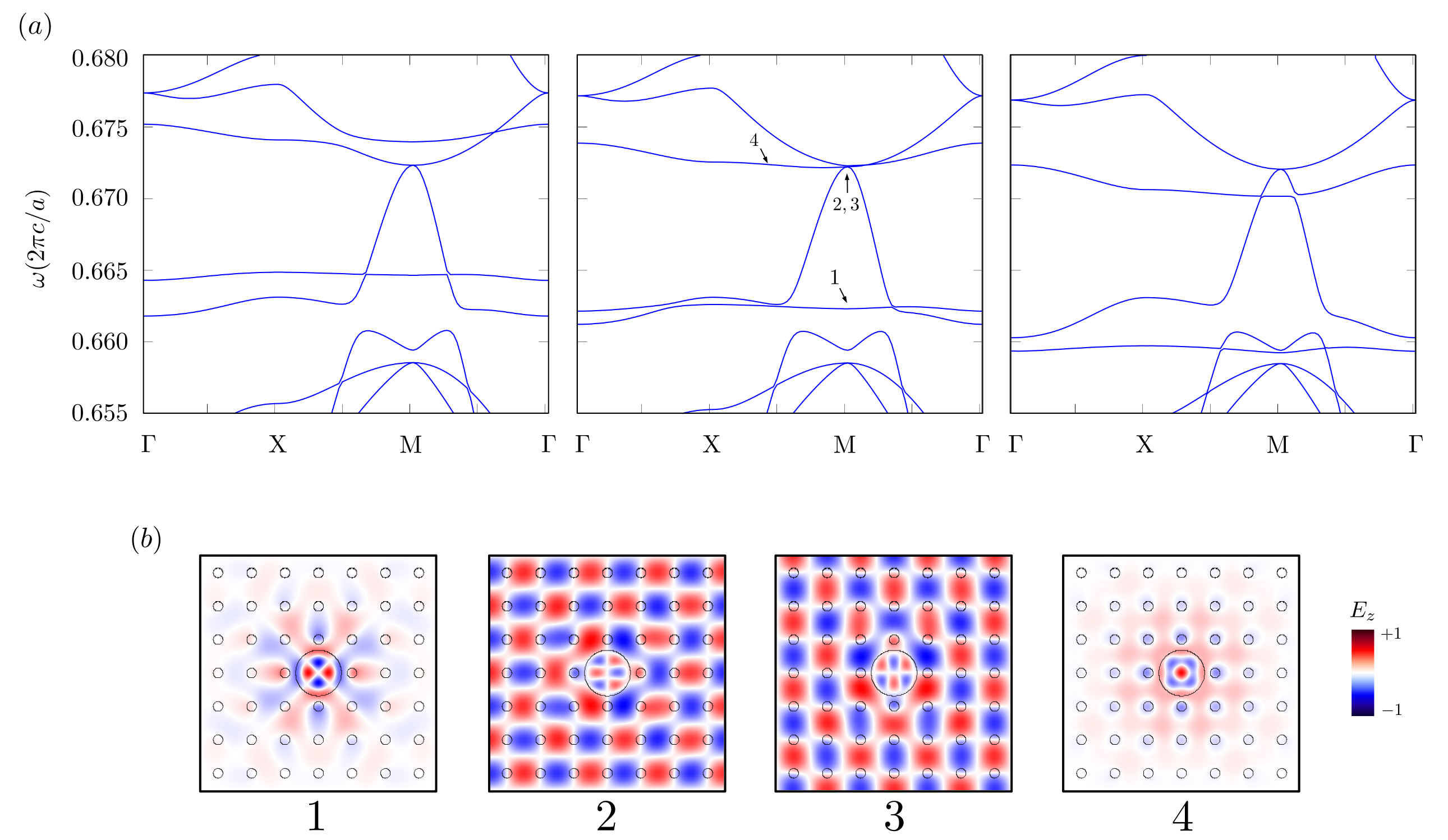}
\caption{(a) The band structure of a supercell consisting of 7x7 sites with periodic boundary conditions. The defect introduced in the center has radii 0.69a, 0.695a and 0.7a in the three sub-plots (left to right). The middle panel shows that the defect mode can be fine-tuned to be degenerate with the spectrally isolated two-fold degeneracy of the bulk. (b) z-component of $\mathbf{E}$-field of the defect modes labelled 1 and 4 (irreps $B_1$ and $A_1$ respectively) and the two modes of the bulk degeneracy labelled 2 and 3 (irrep E).}
\label{fig:figure2s}
\end{figure}

\bibliography{supplement_ref.bib}